\documentclass[aps,prb,floatfix,twocolumn,showpacs,reprint,superscriptaddress]{revtex4-1}
\usepackage{amsfonts}
\usepackage{mathrsfs}
\usepackage{amsmath}
\usepackage{color}
\usepackage{graphicx}
\usepackage{bm}
\usepackage{amssymb}
\usepackage{xspace}
\usepackage{epstopdf}
\usepackage{dcolumn}
\usepackage{longtable}
\usepackage{multirow}
\usepackage[colorlinks=true, letterpaper=true, pdfstartview=FitV, linkcolor=blue, citecolor=blue, urlcolor=blue]{hyperref}
\begin{document}

\preprint{APS/123-QED}

\title{Geometrical Effects in Orbital Magnetic Susceptibility}

\author{Yang Gao}
\affiliation{Department of Physics, The University of Texas at Austin, Austin, Texas 78712, USA}

\author{Shengyuan A. Yang}
\affiliation{Research Laboratory for Quantum Materials and EPD Pillar, Singapore University of Technology and Design, Singapore 487372, Singapore}

\author{Qian Niu}
\affiliation{Department of Physics, The University of Texas at Austin, Austin, Texas 78712, USA}
\affiliation{International Center for Quantum Materials, Peking University, Beijing 100871, China}

\date{\today}

\begin{abstract}
Within the wave-packet semiclassical approach, the Bloch electron energy is derived to second order
in the magnetic field and classified into gauge-invariant terms with clear physical meaning, yielding a fresh
understanding of the complex behavior of orbital magnetic susceptibility. The Berry curvature and quantum metric of the Bloch states
give rise to a geometrical magnetic susceptibility, which can be dominant when bands are filled up to a small energy gap.
There is also an energy polarization term, which can compete with the Peierls-Landau and Pauli magnetism on a Fermi surface.
All these, and an additional Langevin susceptibility, can be calculated from each single band, leaving the Van Vleck susceptibility
as the only term truly from interband coupling.
\end{abstract}


\maketitle

\section{Introduction}
The intrinsic geometrical properties of the Bloch band is of great importance in solid state physics. To zeroth order in electromagnetic fields, the effective mass tensor reflects the curvature of band dispersions, which describes the low energy behaviour near band extrema and enters into the carrier density of states and various transport properties.\cite{Mermin1976} To first order, an accurate description of Bloch electron dynamics not only requires the knowledge of the band dispersion, but also the Berry curvature and the orbital magnetic moment
as functions of the crystal momentum.\cite{Xiao2010}  The Berry curvature reflects the intrinsic geometry of the Bloch state fiber bundle, and its importance has been exemplified in the study of anomalous Hall effects of charge and heat and in the
investigation of orbital magnetization.\cite{Nagaosa2010,Onoda2006, Kagan2008, Zhang2010, Qin2012,Shi2007}

How does the intrinsic geometry of Bloch bands affect the second order response to electromagnetic fields?  Are there additional
geometrical quantities emerge in the orbital magnetic susceptibility?  In this work, we
present an exhaustive analysis of the electron wave-packet energy and orbital magnetic susceptibility by making a gauge invariant classification. We are able to identify a geometrical contributions from the Fermi sea, which we call the geometrical susceptibility, in the sense that it involves geometrical quantities including the Berry curvature and the quantum metric. The geometrical susceptibility is a novel mechanism for orbital magnetic susceptibility, which provides the dominant diamagnetic response around the band gaps, and is especially important in strongly spin/pseudospin-orbit coupled systems such as topological insulators and 2D semimetals.  We propose that by introducing particle-hole symmetry breaking, it is generally possible to enhance the orbital paramagnetism.

Moreover, we derive a novel Fermi surface conribution, arising from the enegy polarization in the Brillouin zone, and competing with Pauli and Peierls-Landau magnetism. To our delight, these Fermi surface contributions, together with a Langevin-like magnetic susceptibility and the geometrical susceptibility, can be calculated based on Bloch states inside a single Bloch band, and the only interband contribution is the Van Vleck paramagnetic susceptibility.
The various terms can dominate over different energy range of the Bloch band. The above understanding of the orbital magnetic susceptibility is under the assumption of the minimal coupling, in which case the magnetic field modifies the Hamiltonian only through the magnetic vector potential. We will also show that our theory can be easily extended beyond the minimal coupling assumption, and be used to describe the effect such as the Zeeman energy from spins.

There are many other systematic studies of the orbital magnetic susceptibility, mostly based on the direct perturbation technique\cite{Adams1953,Hebborn1960,Kohn1955,Roth1962,Blount1962, Wannier1964,Misra1972} or Green\rq{}s function formalism.\cite{Fuku1971,vignale1991,Koshino2007,Schober2012,Santos2011, Raoux2014, Raoux2014b} However, our result has two advantages: (1) it consists of only gauge-invariant terms, which are independent of the phase choice of the Bloch states; (2) it allows simple physical interpretations of each term and hence the precise classification of geometrical and interband effects.

Our paper is organized as follows. In Section II, we demonstrate that the first order correction to the wave-packet has clear physical interpretations and can be understood by two new concepts, i.e. the vertical and horizontal mixings to the unperturbed Bloch states. In Section III, with first order correction to the wave-packet, we derive the wave-packet energy up to second order in magnetic field; in Section IV, we derive the general formula of the orbital susceptibility. By comparing with the orbital susceptibility for atomic systems and free particles we identify two new mechanisms : the geometrical susceptiblity and the energy polarization susceptibility; in Section V and VI we provide two examples to demonstrate the importance of the geometrical effect in orbital magnetic susceptibility and how various contributions compete with each other; in Section VII, we discuss how to extend our theory beyond the minimal coupling assumption.

\section{Vertical and horizontal mixing}

In the first order semiclassical theory, the wave-packet is the superposition of unperturbed local Bloch states, i.e. $e^{i\bm q\cdot \bm p}|u_0 (\bm p+{1\over 2}\bm B\times \bm r_c)\rangle$ where $\bm r_c$ is the center of mass position of the wave-packet. The wave-packet is also assumed to be peaked around some momentum $\bm p_c$ in the Brillouin zone. Then the effective Lagrangian can be calculated for this wave-packet and expressed only in terms of its center of mass position $\bm r_c$, the gauge-invariant crystal momentum $\bm k_c=\bm p_c+{1\over 2}\bm B\times \bm r_c$, and their time derivatives. From the Euler-Lagrange equations of motion, one finds that the dynamics of $\bm r_c$ and $\bm k_c$ contains two geometrical corrections: the orbital magnetic moment $\bm m$ that contributes a Zeeman energy, and the Berry curvature $\bm{\mathit\Omega}$ that modifies the dynamical structure.\cite{Xiao2010}

\begin{figure}[b]
\setlength{\abovecaptionskip}{1pt}
\setlength{\belowcaptionskip}{1pt}
\scalebox{0.6}{\includegraphics*{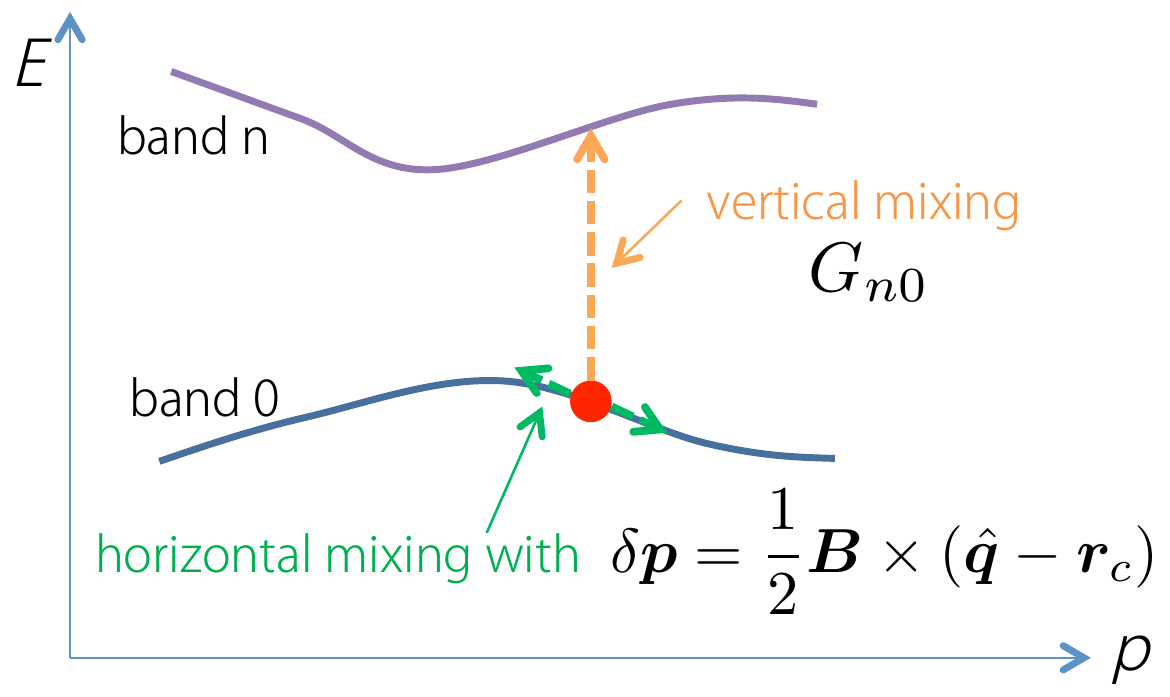}}
\caption{(color online) Schematic figure showing the horizontal mixing and the vertical mixing of Bloch states.}
\label{geometry}
\end{figure}

However, in the second order theory, the local Bloch states and hence the wave-packet should be corrected up to first order. These corrections have two origins: (1) the non-adiabatic correction due to the fact that the local Hamiltonian itself is implicitly time-dependent through the semiclassical evolution of the center of mass position $\bm r_c$; (2) the first order gradient correction to the local Hamiltonian. They will modify the local Bloch state in two ways (see Fig.(\ref{geometry})): (1) they can mix the Bloch states from other bands but at the same point in the Brillouin zone, which we call the vertical mixing; (2) they can also mix the Bloch states inside the same band but from different points in the Brillouin zone.

To derive the first order corrections to the Bloch states, note that the wave-packet is assumed to be localized in real space near the center of mass position $\bm r_c$, we expand the exact Hamiltonian in weak magnetic field with respect to the deviation $\hat{\bm q}-\bm r_c$ from $\bm r_c$ (assume minimal coupling and set $e=\hbar=1$ for simplicity):\cite{Xiao2010,Yang2014} ${\hat H}=\hat{H}_c+\hat{H}^\prime+\hat{H}^{\prime\prime}+\cdots$. Here $\hat{H}_c(\hat{\bm p},\hat{\bm q})=\hat{H}_0(\hat{ \bm p}+{1\over 2}\bm B\times \bm r_c,\hat{\bm q})$ is the local Hamiltonian by taking the magnetic vector potential at $\bm r_c$, where $\hat{H}_0$ is the Hamiltonian without external fields, and $\hat{\bm p}$, $\hat{\bm q}$ are the momentum and position operators. $\hat{H}^\prime$ is the first order gradient correction to $\hat{H}_c$: $\hat{H}^\prime=-{1\over 2}\bm B\cdot [\hat{\bm V}\times (\hat{\bm q}-\bm r
 _c)]$, where $\hat{\bm V}=-i[\hat{\bm q},\hat{H}_0]$ is the velocity operator. $\hat{H}^{\prime\prime}={1\over 8}\mathit\Gamma_{ij} [\bm B\times (\hat{\bm q}-\bm r_c)]_i [\bm B\times (\hat{\bm q}-\bm r_c)]_j$ is the second order perturbation to $\hat{H}_c$, where $\mathit\Gamma_{ij}=\partial_{ p_i p_j} \hat{H}_0$ is the Hessian matrix. For nonrelativistic Pauli and Schr$\ddot{\rm o}$dinger Hamiltonians, $\mathit \Gamma_{ij}=\delta_{ij}/m$, where $m$ is the bare electron mass. For the relativistic Dirac Hamiltonian, $\mathit\Gamma_{ij}=0$. For calculating the orbital magnetic susceptibility, corrections to $\hat{H}_c$ beyond second order will not contribute, and hence they are not included in the following discussion.

By using the above gradient expansion of the Hamiltonian in the time-dependent Schr$\ddot{\rm o}$dinger equation for the wave-packet, we derive the modification to the periodic part of the Bloch states $|u_0(\bm p+{1\over 2}\bm B\times \bm r_c)\rangle$:\cite{Yang2014}
\begin{equation}\label{bloch}
|\tilde{u}\rangle=\lambda |u_0(\bm p+{1\over 2}\bm B\times \hat{\bm q})\rangle+\sum_{n\neq 0}{G_{n0}\over \varepsilon_0-\varepsilon_n} |u_n(\bm p+{1\over 2}\bm B\times {\bm r}_c)\rangle\,,
\end{equation}
where $\lambda$ is a normalization factor to ensure $\langle \tilde{u}|\tilde{u}\rangle=1$ to first order, $G_{n0}=-\bm B\cdot \mathcal{M}_{n0}$, $\mathcal{M}_{n0}={1\over 2}(\sum_{m\neq 0}\bm V_{nm}\times \bm A_{m0}+\bm v_0\times \bm A_{n0})$,  $\bm A_{n0}=\langle u_n|i\bm \partial |u_0\rangle$ is the interband Berry connection, $\bm V_{nm}=\langle u_n |\hat{\bm V}|u_m\rangle$ is the velocity matrix element, and $\bm v_0\equiv \bm V_{00}$. The subscripts $n$, $m$, and $0$ are band indices. $\varepsilon$ represents the band energy. The partial derivative $\bm \partial$ is with respect to the crystal momentum.

From this modified Bloch states in Eq.(\ref{bloch}), the influence of the magnetic field is quite heuristic and preludes the form of the wave-packet energy in Eq.(\ref{energy}).
First consider the second term in Eq.(\ref{bloch}). It contains local Bloch states at the same momentum from other bands ($n\neq 0$) and preserves the lattice translational symmetry. The essential quantity $G_{n0}$ is gauge-invariant and its form is quite suggestive, which are couplings between the magnetic field and interband matrix elements of magnetic dipole moment $\mathcal{M}_{n0}$. Note that the correct expression of $\mathcal{M}_{n0}$ includes the time dependence of $\bm r_c$ in the Bloch states, which is in fact the nonadiabatic effect mentioned previously.\cite{Yang2014} The remaining part in $\bm B\cdot \mathcal{M}_{n0}$ is from the interband part ($\bm A_{n0}$) of the postion operator $\hat{\bm q}$ in $\hat{H}^\prime$ in the Bloch representation. We call this correction with $G_{n0}$ the vertical mixing since it mixes Bloch states from different bands at the same $k$-point in the Brillouin zone (as illustrated in Fig.(\ref{geometry})).

On the contrary, the first term in Eq.(\ref{bloch}) only contains the Bloch state inside the same band (band 0). Compared with the original Bloch state $|u_0(\bm p+{1\over 2}\bm B\times \bm r_c)\rangle$, its crystal momentum is shifted to $\bm p+{1\over 2}\bm B\times \hat{\bm q}$, hence breaking the lattice translational symmetry. An interesting fact is that this term can be obtained if we initially take the position operator in the vector potential in the exact Hamiltonian $\hat{H}$ as a c-number and later recover it in $|u_0\rangle$ as an operator. This shift of momentum is due to the intraband part of the postion operator $\hat{\bm q}$ in $\hat{H}^\prime$ in the Bloch representation. We emphasize that such type of correction without lattice trasnlational symmetry must appear, since the exact Hamiltonian $\hat{H}$ has an argument $\bm p+{1\over 2}\bm B\times \hat{\bm q}$ and does not generally respect such symmetry. Heuristically, the correct single band wave-packet should be constructed from the eigenstates of the exact Hamiltonian which does not respect the lattice translational symmetry in general. 
However, since the susceptibility is a second order thermodynamic response, we can approximate $\hat{H}$ by $\hat{H}_c$, with gradient corrections up to second order. Unlike $\hat{H}$, $\hat{H}_c$ is periodic with a well-defined Brillouin zone. The influcence of the actual lattice translational symetry breaking thus manifests as a first order correction to the local Bloch states in the form of the momentum shift.

Remarkably, this correction to $|u_0\rangle$ can be rewritten simply in terms of the shift of momentum $\delta \bm p={1\over 2}\bm B\times (\hat{\bm q}-\bm r_c)$, and reads $\delta \bm p \cdot \hat{\bm D}|u_0\rangle$, where $\hat{\bm D}=\bm\partial+i\bm a_0$ is the gauge-covariant derivative acting on the Bloch states and $\bm a_{0}=\langle u_0|i\bm \partial |u_0\rangle$ is the intraband Berry connection. We will see that the momentum shift $\delta \bm p$ enters the description of electrons\rq{} orbital motion through the band geometrical properties, yielding crucial geometrical corrections as discussed later. Since the correction from $\delta \bm p$ mixes Bloch states at neighbouring $k$-points in the same band, we call it the horizontal mixing.  

To sum up, the magnetic field affects the local Bloch states $|u_0\rangle$ in two ways: (i) it vertically mixes $|u_0\rangle$ with the Bloch states $|u_n\rangle$ from other bands; (ii) it also horizontally shifts the Bloch states along the path $\delta \bm p$ according to the affine connection (Berry connection) $\bm a_0$ in the Brillouin zone. We illustrate the two types of mixing of Bloch states schematically in Fig.\ref{geometry}.

\section{Wave-packet energy up to second order correction}

With the modified Bloch states derived in the previous section, we can construct the corrected wave-packet $|\Psi\rangle$ up to first order and derive the corresponding effective Lagrangian up to second order.\cite{Yang2014} From this Lagrangian, the wave-packet energy is obtained, which is the crucial quantity for evaluating the orbital magnetic susceptibility. By expressing the wave-packet energy in a gauge-invariant form, the physical meaning of various contributions is interpreted. The concept of vertical and horizontal mixings is important to understand the wave-packet energy.

From Eq.(\ref{bloch}), the corrected wave-packet up to first order reads:
\begin{equation} \label{wavepacket}
|\Psi\rangle=\int d\bm p e^{i\bm p\cdot \bm q} \left( C_0(\bm p)|u_0\rangle+\sum_{n\neq 0} C_n(\bm p) |u_n\rangle\right)\,,
\end{equation}
where $C_0$ is a coefficient peaked at $\bm p_c$: $ |C_0|^2=\delta(\bm p-\bm p_c)$, $C_n$ is connected to $C_0$ based on Eq.(\ref{bloch}):
\begin{equation}\label{Cn}
C_n=-i{1\over 2}[\bm B\times (i\bm \partial+\bm a_0-\bm r_c)C_0]\cdot \bm A_{n0}-{G_{n0}\over \varepsilon_0-\varepsilon_n} C_0\,.
\end{equation}
In Eq.(\ref{Cn}), the first term is derived from the horizontal mixing by integration by parts, and the second term is from the vertical mixing in Eq.(\ref{bloch}).

The effective Lagrangian for the wave-packet reads $L=\langle\Psi|i\partial_t-\hat{H}_c-\hat{H}^\prime-\hat{H}^{\prime\prime}|\Psi\rangle$. It can be transformed to a Langrangian for the single band wave-packet: $L=\int C_0^\star(\bm p) i\partial_t C_0(\bm p) d\bm p-\tilde{\varepsilon}$.
We then calculate the wave-packet energy up to second order as follows: $\tilde{\varepsilon}=\langle \Psi |\hat{H}_c+\hat{H}^\prime+\hat{H}^{\prime\prime}|\Psi\rangle+\delta L_{dyn}$, where $\delta L_{dyn}=\int C_0^\star(\bm p) i\partial_t C_0(\bm p) d\bm p-\langle \Psi|i\partial_t |\Psi\rangle$ is the contribution from the dynamical part of the Lagrangian.
The wave-packet energy corrected up to second order reads (see Appendix A for details):
\begin{align}\label{energy}
\tilde{\varepsilon}=\,&\varepsilon_0 -\bm B\cdot \bm m\notag\\
&+{1\over 4} (\bm B\cdot \bm{\mathit\Omega}) (\bm B\cdot \bm m)-{1\over 8} \epsilon_{s i k} \epsilon_{t j \ell}B_{s} B_{t}g_{ij} \alpha_{k\ell} \notag\\
& -\bm B\cdot (\bm a_0^\prime \times \bm v_0)+\nabla\cdot\bm P_\text{E}\notag\\
&+\sum_{n\neq 0}{G_{0n} G_{n0}\over \varepsilon_{0}-\varepsilon_{n}}+{1\over 8m} (B^2 g_{ii}-B_i g_{ij} B_j).
\end{align}
Here $\bm{\mathit\Omega}=-{\rm Im}\langle \bm \partial u_0|\times |\bm \partial u_0\rangle$ is the Berry curvature, $\bm m=-{1\over 2} {\rm Im}\langle \bm \partial u_0|\times (\varepsilon_0-\hat{H}_c) |\bm \partial u_0\rangle$ is the orbital magnetic moment, $g_{ij}={\rm Re}\langle \partial_i u_0|\partial_j u_0\rangle-(a_0)_i (a_0)_j$ is the quantum metric of $k$-space,\cite{anandan1990, Neupert2013} $\alpha_{k\ell}=\partial_{k\ell}\varepsilon_0$ is the inverse of effective mass tensor, $\bm a_0^\prime=\sum_{n\neq 0} [G_{0n}\bm A_{n0}/(\varepsilon_0-\varepsilon_n)]+{1\over 4}\partial_i[(\bm B\times \bm A_{0n})_i\bm A_{n0}]+c.c.$ is the field-induced positional shift of the wave-packet center (its second term is a geometrical quantity from the horizontal mixing and is proportional to the Christoffel symbol in $k$-space\cite{Yang2014}), and $\bm P_\text{E}=(1/4)[
\langle(\bm B\times\hat{\bm D})u_0|(\hat{\bm V}+\bm v_0)\cdot|(\bm B\times\hat{\bm D})u_0\rangle+c.c.]$ is a single band quantity representing the energy polarization density in $k$-space. Indices
$i$, $j$, $k$, $\ell$, $s$ and $t$ refer to Cartesian coordinates and repeated indices are summed over. $\epsilon_{sik}$ is the totally antisymmetric tensor in three dimension. For the last term in Eq.(\ref{energy}), we choose $\mathit{\Gamma}_{ij}=\delta_{ij}/m$ (for Pauli and Schr$\ddot{\rm o}$dinger Hamiltonians) for simplicity and a more general formula is given in Appendix A. All physical quantities in Eq.(\ref{energy}) should be understood as functions of the gauge-invariant crystal momentum $\bm k_c$, and the partial derivatives are with respect to $\bm k_c$.

In Eq.(\ref{energy}), various terms are gauge-invariant and can be characterized by their geometrical and physical meaning.
The two terms in the first line of Eq.(\ref{energy}) are the band energy plus the magnetic dipolar energy, which is the result obtained in the first order semiclassical theory. 

The two terms in the second line are the geometrical energies, in the sense that they consist of single band geometrical quantities, i.e. the Berry curvature and the quantum metric. On the Brillouin zone, the Hilbert space with single band Bloch states $|u_0\rangle$ forms a fiber bundle, whose curvature is characterized by the Berry curvature $\bm{\mathit\Omega}$,\cite{Xiao2010} and the distance in which is captured by the quantum metric.\cite{Neupert2013} For the remaining two quantities, $\alpha_{kl}$ depends only on the band dispersion and the magnetic moment $\bm m$ is a single band quantity.
It is interesting to note that $\alpha_{ij}$ and $\bm m$ actually form a conjugate pair: they are proportional to the real and the imaginary part of $\delta_{ij}/m+2\langle \partial_i u_0|(\varepsilon_0-\hat{H}_c)|\partial_j u_0\rangle$, respectively. So are the quantum metric and the Berry curvature, with respect to the quantity $\langle \partial_i u_0|\partial_j u_0\rangle-(a_0)_i(a_0)_j$. Thus the less obvious meaning of $g_{ij}$ and $\bm m$ can be understood from their well studied conjugate partners $\bm{\mathit\Omega}$ and $\alpha_{ij}$.

The first term in the third line of Eq.(\ref{energy}) is a real space polarization energy. The magnetic field shifts the wave-packet center by $\bm a_0^\prime$,\cite{Yang2014} hence modifying the magnetic dipole moment and the wave-packet energy. The next term is a $k$-space polarization energy. This can be understood by noticing that the the momentum shift $\delta \bm p$ gives rise to a second order energy polarization in $k$-space, $(1/2)(\hat{H}^\prime \delta \bm p+c.c.)$. Similar to the relation between electric polarization and charge, the divergence of such energy polarization yields a local energy correction. We find that this term is a single band quantity, and is related to the quadrupole moments of the velocity operator (see Appendix B for details).

In the fourth line of Eq.(\ref{energy}), the first term is a standard second order perturbation energy through virtual interband transitions. The last term in Eq.(\ref{energy}) is from the perturbation of $\hat{H}^{\prime\prime}$. Note that this term vanishes for the Dirac Hamiltonian, and for the nonrelativistic Pauli and Schr$\ddot{\rm o}$dinger Hamiltonians, it comes with the bare electron mass $m$.

The above geometrical and physical meanings of the second order wave packet energy are also suggested by the vertical and horizontal mixings in the wave packet. Such two types of corrections in $|\Psi\rangle$ originates from the Bloch represention of $\hat{H}^\prime$. Therefore, they enter the wave packet energy in Eq.(\ref{energy}) through both $\hat{H}^\prime$ and $|\Psi\rangle$. If two horizontal mixings are combined to yield a second order energy, only the neighbourhood in the Brillouin zone is involved, and we should obtain a purely geometrical contribution as in the second line in Eq.(\ref{energy}). On the contrary, if two vertical mixings are combined, then virtual interband transition is involved, and we obtain an interband effect as the first term in the fourth line of Eq.(\ref{energy}). If the horizontal and vertical mixing  are combined together, we obtain the $k$-space polarization energy, which is a single band but not necessarily geometryical quantity.

\section{Orbital Magnetic Susceptibility}

The various second order energy corrections in Eq.(\ref{energy}) are indispensable for the evaluation of the orbital magnetic susceptibility. The general approach is to evaluate the thermodynamic grand potential $G=\text{Tr}[g(\hat{ H})]$, where $g(\varepsilon)=-k_BT\ln(1+{\rm exp}[(\mu-\varepsilon)/k_B T])$, $T$ is the temperature and $\mu$ is the chemical potential. Then the magnetic susceptiblity is simply the second order derivatives of this grand potential with respect to the magnetic field. By comparing with the susceptibility in atomic systems and free particles, the physical meaning of various terms in the orbital susceptibility is interpreted. Various mechamisms are more clear when we rewrite the free energy in the Wannier function representation and take the atomic insulator limit. We find that in solids, there is a novel and purely geometrical contribution which depends solely on geometrical properties of Bloch bands in the Brillouin zone, and we call it the geometrical susceptibility.

Under external magnetic field, the semiclassical limit of the grand potential written in terms of physical variables is given by:\cite{Blount1962}
\begin{equation}\label{fe}
G=V\int_\text{BZ}[\mathcal{D}g(\tilde{\varepsilon})+g_\text{L}]{d^3 k_c\over 8\pi^3}.
\end{equation}
Here $V$ is the system volume and $\mathcal{D}=1+\bm B\cdot (\bm{\mathit\Omega}+\nabla\times \bm a_0^\prime)$ is the modified density of states.\cite{Yang2014} The first term in the bracket of Eq.(\ref{fe}) is from the semiclassical grand potential density with second order energy correction, which yields the semiclassical free energy. The second term $g_\text{L}$ is the Peierls-Landau magnetic energy:
$g_\text{L}=-(f^\prime/48)B_s B_t \epsilon_{s i k}\epsilon_{t j \ell} \alpha_{i j} \alpha_{k\ell}\,,$
where $f^\prime$ is the energy derivative of the Fermi distribution function $f$. For isotropic bands, the effective mass tensor $\alpha$ takes a diagonal form, and $g_\text{L}$ reduces to its familiar form.\cite{Mermin1976} This term originates from the discreteness of the Landau levels, and appears when we transform the free energy from the quantum version to its semiclassical limit.\cite{Blount1962}

We combine Eq.(\ref{energy}) with Eq.(\ref{fe}) and expand the free energy to second order: $G=V\int_\text{BZ} (g_0+g^\prime+g^{\prime\prime}) d^3k_c/(8\pi^3)$. At zeroth order, $g_0=g(\varepsilon_0)$. At first order, $g^\prime=-\bm B\cdot \bm m f+\bm B\cdot \bm{\mathit\Omega} g_0$, which yields the same magnetization as in Ref.\onlinecite{Shi2007}. The second order $g^{\prime\prime}$ is required for the magnetic susceptibility $\chi_{ij}=-(1/V)(\partial^2 G/\partial B_i\partial B_j)_{\mu,T,V}$, and reads (for simplicity we take $\mathit{\Gamma}_{ij}=\delta_{ij}/m$)
\begin{align}\label{sus}
g^{\prime\prime}&=g_\text{L}+{f^\prime\over 2}(\bm B\cdot \bm m)^2-{f^\prime\over 4} \bm v_0\cdot \bm P_\text{E}\notag\\
&+f{G_{0n}G_{n0}\over \varepsilon_0-\varepsilon_n} +{f\over 8m} (B^2 g_{ii}-B_i g_{ij} B_j)\notag\\
&-{3f\over 4}(\bm B\cdot \bm{\mathit\Omega}) (\bm B\cdot \bm m)-{f\over 8} \epsilon_{s i k} \epsilon_{t j \ell}B_{s} B_{t}g_{ij}\alpha_{k\ell}.
\end{align}
Magnetisms in the first line of Eq.(\ref{sus}) are contributions from the Fermi surface. The first two contributions are the Peierls-Landau magnetism, and the Pauli paramagnetism for the orbital moment $\bm m$. In solids, the Peierls-Landau magnetism can be paramagnetic, especially near the band saddle points.\cite{vignale1991} These two contributions are prominant around singular points where the density of states diverges. For example, around the saddle point, Peierls-Landau magnetism dominates in general, \cite{vignale1991} and for the spin-1 continuum model, the Pauli paramagnetism contributes to a paramagnetic peak at $\mu=0$ where the flat band emerges as explained in Sect.VI. The third term is due to the $k$-space energy polarization in Eq.(\ref{energy}), and is first identified here. Similar to the Pauli and Peierls-Landau magnetism, it also involves only single band quantities.This term generally compete with the Pauli orbital and Peierls-Landau magnetisms as illustrated in Sect.V, except at the band maxima or minima where $|\bm v_0|$ vanishes. 

The other terms in Eq.(\ref{sus}) are Fermi sea contributions. The first term in the second line yields the Van Vleck susceptibility originated from the vertical mixing energy in Eq.(\ref{energy}). It is the only interband contribution to the orbital magnetic susceptibility. It is always paramagnetic after summing over all the occupied bands, similar to the Van Vleck paramagnetic susceptibility in atomic systems. The second one yields a Langevin-like magnetic susceptibility from the last term in Eq.(\ref{energy}). It can be expressed in a compact form using the quantum metric $g_{ij}$, which describes the intrinsic fluctuation of position-position operators ($\hat{q}_i\hat{q}_j$) in the Bloch representation: $g_{ij}={\rm Re}[(A_i)_{0n}(A_j)_{n0}]$ (see Appendix B for details). For Pauli or Schr$\ddot{\rm o}$dinger Hamiltonian with constant mass, this term yields diamagnetic response along directions that diagonalize $g_{ij}$. Its expression will change for effective Hamiltonians with a general Hessian matrix $\mathit\Gamma_{ij}$ as given in Appendix A. For Dirac Hamiltonian, the Lagnevin-like magnetic susceptibility vanishes.

Magnetic free energies in the third line in Eq.(\ref{sus}) have no analogs as in atomic physics or for free particles, and are first identified here. We call the susceptibility from these two terms the geometrical magnetic susceptibility, because they are due to the geometrical energies in Eq.(\ref{energy}) from the horizontal mixing of Bloch states and the geometrical correction to the density of states in Eq.(\ref{fe}). Notice that the first term consists of the Berry curvature, it is important when the band structure contains monopole or other nontrivial topological structures. For example, for two band systems with the particle-hole symmetry, geometrical magnetic susceptibility always yields a diamagnetic susceptibility and is a prominant or even dominant contribution in the band gap.

In the limit of atomic insulators where the hopping between lattice sites is suppressed, these contributions in the second line in Eq.(\ref{sus}) will reduce to the familiar Van Vleck paramagnetic susceptibility and Langevin diamagnetic susceptibility in atomic physics. To demonstrate this, we use the Wannier function representation. Notice that the periodic part of the Bloch function can be expressed in terms of the Wannier function:
\begin{equation}\label{wannier}
|u_n(\bm k_c)\rangle={1\over \sqrt{N}} \sum_{\bm R} e^{-i\bm k_c\cdot (\bm r-\bm R)} |W_n(\bm r,\bm R)\rangle\,,
\end{equation}
where $n$ is the band index and $|W_n(\bm r,\bm R)\rangle$ is the Wannier function localized at $\bm R$ for band $n$. With the help of Eq.(\ref{wannier}), we can express the Van Vleck paramagnetic susceptibility, Langevin magnetic susceptibility and geometrical magnetic susceptibility in terms of Wannier functions. To simplify the discussion, we consider an insulator (with each band either completely filled or empty) with a single atom at each lattice site and then suppress the inter-lattice-site hopping (see Appendix C for details). The Van Vleck term reads:
\begin{align}
&\quad\int {d^3k_c\over 8\pi^3} \sum_{n}f(\varepsilon_0(\bm k_c)){G_{0n}G_{n0}\over \varepsilon_0-\varepsilon_n}\notag\\
&= {N\over V}\sum_n{f(\varepsilon_0)\over \varepsilon_0-\varepsilon_n}\big | \langle W_n(\bm R)|{1\over 2}\bm B\cdot[\hat{\bm V}\times (\bm r-\bm R)] |W_0(\bm R)\rangle|^2\,,
\end{align}
where $N$ is the total number of atoms and $V$ is the sample volume. Notice that this is the same as the Van Vleck paramagnetic susceptibility in atomic systems for the energy level $\varepsilon_0$.
The Langevin term reads:
\begin{align}
&\quad\int {d^3k_c\over 8\pi^3} \sum_{n}f(\varepsilon_0(\bm k_c)){f\over 8m} (B^2 g_{ii}-B_i g_{ij} B_j)\notag\\
&={N\over V}{f(\varepsilon_0)\over 8m} \langle W_0(\bm R)||\bm B\times (\bm r-\bm R)|^2|W_0(\bm R)\rangle\,,
\end{align}
which is the familiar Langevin diamagnetic susceptibilityin atomic systems. Furthermore, the remaining Fermi sea contribution to the orbital susceptibility in Eq.(\ref{sus}), i.e. the geometrical magnetic susceptibility, vanishes in the atomic insulator limit by similar derivations. Therefore, our formula indeed reduces to the correct result in the atomic insulator limit.

To conclude the discussion of the orbital susceptibility in the Wannier funtion representation, we comment that  the Fermi sea contribution to the orbital susceptibility is due to two types of effects, i.e. intra-lattice-site transition similar to that in the atomic physics and inter-lattice-site hopping which is unique in crystalline solids. Our classification in Eq.(\ref{sus}) thus provide a reasonable extrapolation of the orbital susceptibility from atomic systems to crystalline solids: on one hand, the Van Vleck paramagnetic susceptibility and Langevin magnetic susceptibility reduce to their counterparts in atomic systems and the geometrical magnetic susceptibility vanishes in the limit of atomic insulators; on the other hand, even though they all contain inter-lattice-site hopping contribution, the Van Vleck paramagnetic susceptibility and the Langevin magnetic susceptibility in solids still preserve their essential properties established in the atomic systems, i.e. the Van Vleck paramagnetic susceptibility is always paramagnetic and depends on the energy interval between two electronic states and the Langevin magnetic susceptibility is diamagnetic along directions that diagonalize the quantum metric. Furthermore, we emphasize that the geometrical magnetic susceptibility is indeed unique in crystalline solids and is a novel mechanism of orbital susceptibility that only depends on the geometrical quantities in $k$-space.

In atomic systems, the Van Vleck paramagnetic susceptibility is generally small due to the large separation between electronic levels, although it has a notable exception in the atomic Lanthanide series, where the energy interval between the ground state and the first excited states is small. In solids, different energy levels can be very close, e.g. near the topological transition points. In this case, the Van Vleck paramagnetic susceptibility become large. However, we will show that in the circumstance where the linear band crossing occurs, it is another contribution, i.e. the geometrical contribution, that dominates over other orbital susceptibilities. The Langevin magnetic susceptibility in atomic systems is also small and only important for close shell atoms. Likewise, in solids the Langevin magnetic susceptibility is usually discussed for insulators where electrons are localized. However, our theory suggests that it is connected to the intrinsic expansion of the localized wave-packet, i.e. the quantum metric (see Appendix B for details). Therefore, for both insulators and metals, the Langevin mangetism is well defined and  can be sizable as illustrated in Sect.V. It can contribute to the prominant paramagnetic plateau as discussed in Ref.\onlinecite{Raoux2014}. It can even dominate over other susceptibilities as in the continuum model of the double layer graphene.

Eq.(\ref{sus}) and Eq.(\ref{energy}) are the main results of this work. By developing a gauge-invariant second order semiclassical theory, we successfully obtain the complete orbital magnetic susceptibility in a compact form, with each term gauge-invariant and having a clear physical meaning. Of all the terms in Eq.(\ref{sus}), only the Van Vleck contributions involve interband processes, while other terms are single band properties. Particularly, we are able to cast these single band terms (except for energy polarization and the Langevin-like term in general case) in a form only involving the single band quantities $\alpha_{ij}$ and $\bm m$, and geometrical quantities $\bm{\mathit\Omega}$ and $g_{ij}$, clearly demonstrating their intrinsic geometrical identity.

\section{Example I: Honeycomb lattice model}
As a concrete example to show how various terms contribute to the total magnetic susceptibility and the importance of the geometrical magnetic susceptibility, we first consider the following tight-binding model defined on a honeycomb lattice:\cite{NG2009}
\begin{align}\label{honey}
\hat{H}=&-t\sum_{\langle i,j\rangle}c_{i}^\dagger c_{j}-t^\prime \sum_{\langle\langle i,j\rangle\rangle} c_{i}^\dagger c_{j}
+\Delta \sum_{i} \xi_i c_{i}^\dagger c_{i}\,,
\end{align}
where $c_i(c_i^\dagger)$ is the electron annihilation (creation) operator on site $i$, the first and the second terms are the nearest-neighbor and second-neighbor hopping terms, the third term is a staggered potential with $\xi_i=\pm 1$ for the two sublattices, and $t$, $t'$ and $\Delta$ are the strengths of the terms. The staggered potential breaks the inversion symmetry and generates a gap of $2\Delta$ in the spectrum. The second-neighbor hopping is introduced for breaking the particle-hole symmetry.

\begin{figure}[b]
\setlength{\abovecaptionskip}{1pt}
\setlength{\belowcaptionskip}{1pt}
\scalebox{0.46}{\includegraphics*{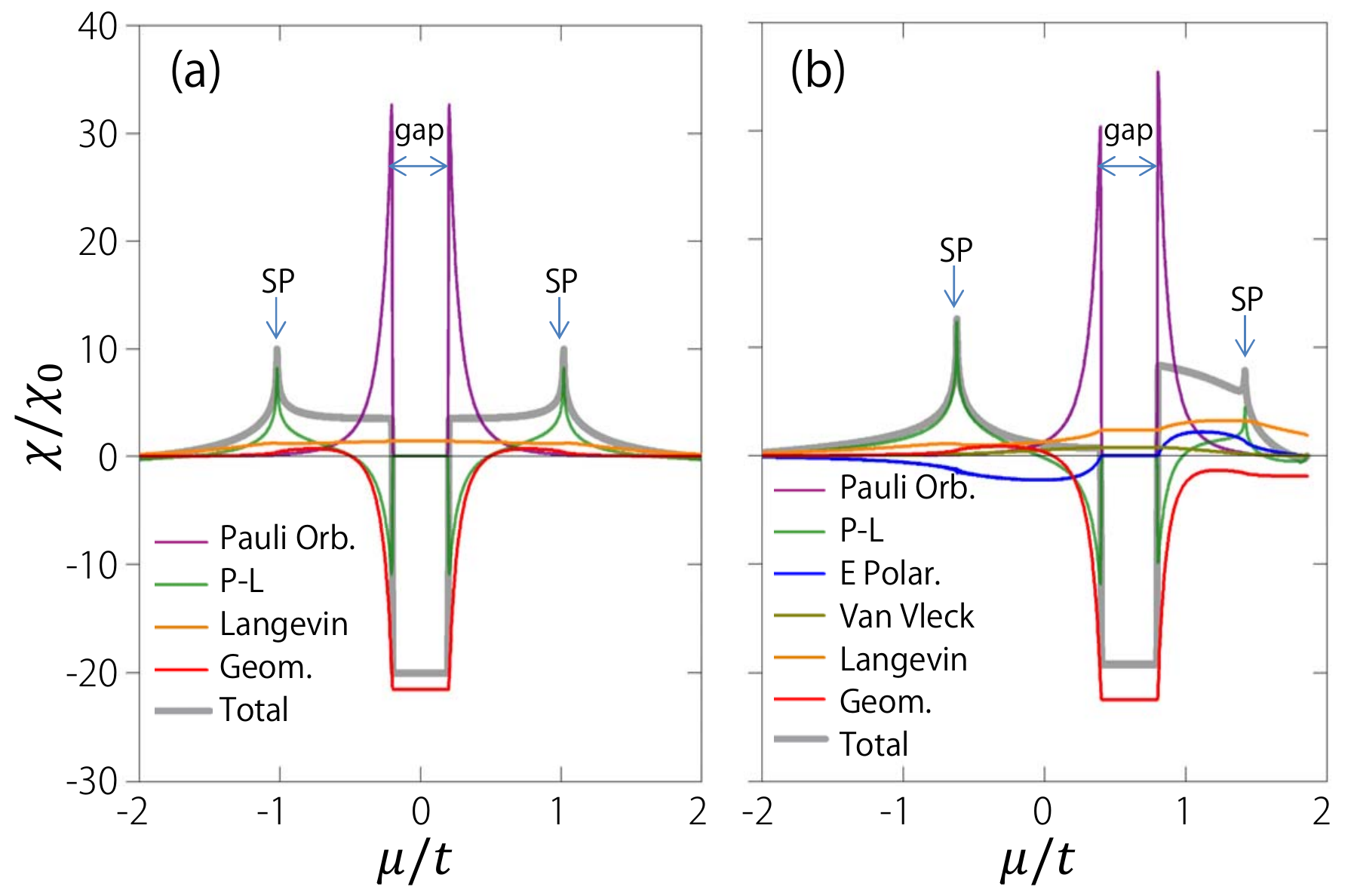}}
\caption{(color online) Orbital magnetic susceptibility for the lattice model (\ref{honey}) as a function of $\mu$. $\chi$ is in units of $\chi_0=e^2a^2t/(4\pi^2\hbar^2)$, $a$ is the bond length. Here $\Delta=0.2t$, (a) $t^\prime=0$ and (b) $t^\prime=0.1t$. Here P-L, E Polar, and SP stand for the Peierls-Landau, energy-polarization, and saddle point, respectively.}
\label{f_sus}
\end{figure}

The various contributions to the orbital magnetic susceptibility are plotted in Fig.\ref{f_sus} with (a) and without (b) particle-hole symmetry.
In the presence of particle-hole symmetry ($t'=0$), the energy polarization and the Van Vleck contributions vanish identically. From Fig.\ref{f_sus}(a), one observes that in the gap the Fermi surface terms vanish and the geometrical magnetic susceptibility dominates, which leads to a large diamagnetic response. The magnitude of the geometrical magnetic susceptibility decreases rapidly away from the gap and it (along with Peierls-Landau term) is compensated largely by the Pauli orbital paramagnetic susceptibility which is peaked at the band edges where $\bm m$ takes its largest value.\cite{Cai2013} Two noticeable paramagnetic peaks are observed around the band saddle points due to the Peierls-Landau contribution, which is a general feature as discussed before.\cite{vignale1991} Further away from the gap region, the susceptibility decreases gradually to zero. Our result of $\chi$ agrees with that from the exact quantum treatment.\cite{Raoux2014}

The physics around gap can be described by the gapped Dirac model $\hat{h}=vk_1 \sigma_1+vk_2\sigma_2+\Delta \sigma_3$ with $v=3at/2$ where $a$ is the nearest-neighbor bond length.\cite{NG2009} This model is widely used in the study of graphene, ${\rm MoS_2}$, topological insulator surfaces and thin films.\cite{NG2009, Xiao2012, Zhang2011, Kane2010} Here $\sigma$'s are the Pauli matrices. Near the band edge ($|\mu|>\Delta$), the three competing magnetic susceptibilities (Pauli $\chi_\text{P}$, Peierls-Landau $\chi_\text{L}$, and geometrical $\chi_\text{Geom}$) read explicitly
\begin{equation}\label{Dirac}
\chi_\text{P}={e^2\Delta^2 v^2\over 8\pi |\mu|^3},\; \chi_\text{L}=-{\chi_\text{P}\over 3},\; \chi_\text{Geom}=-{e^2\Delta^2 v^2\over 12 \pi |\mu|^3}.
\end{equation}
We emphasize that for systems with two valleys connected by time reversal operation, such as graphene or ${\rm MoS_2}$, the geometrical magnetic susceptibilities from the two valleys have the same sign. Note that in this low energy model the total susceptibility vanishes identically outside gap, which seems contradicting to the result in Fig.\ref{f_sus}(a) where one sees a finite paramagnetic plateau. The difference is due to two Fermi sea contributions including the Langevin and a term in geometrical magnetic susceptibility resulting from the nonzero Hessian $\mathit\Gamma_{ij}$ (whereas $\mathit\Gamma_{ij}$ vanishes in the low energy model) which produce an overall shift of $\chi$. This was known as the ``lattice contribution" in previous studies.\cite{Santos2011} This agreement demonstrates the validity of our theory.

We point out that if one starts from the Pauli or Schr$\ddot{\rm o}$dinger Hamiltonian, the Langevin magnetic susceptibility is diagmanetic as discussed previously. But for an effective Hamiltonian as given in Eq.(\ref{honey}), the Langevin magnetic susceptibility can be paramagnetic as shown in Fig.\ref{f_sus}. Therefore,  this paramagnetic plateau is sizable as in Fig.\ref{f_sus} only when these two bands are well separated from other bands such that the tight binding model is an appropriate approximation.

When particle-hole symmetry is broken by a finite $t'$, as shown in Fig.\ref{f_sus}(b), the geometrical magnetic susceptibility still dominates in the gap and one notes that the paramagnetic plateau near gap is suppressed in the valence band while enhanced in the conduction band. Now the energy polarization and the Van Vleck contributions are finite. With current parameters, the Van Vleck paramagnetic susceptibility takes a small value around the gap region, while the energy polarization contribution takes opposite signs between the two bands. The energy polarization term, along with the enhanced Langevin and Pauli terms are the main contributions to a large paramagnetic response between the conduction band edge and the saddle point. This is different from the usual orbital paramagnetic susceptibility resulting from the Peierls-Landau contribution.\cite{vignale1991}  In fact, the contribution from Peierls-Landau is less important even 
in the region near and above the saddle point in conduction band. The total susceptibility there is more affected by the competition between geometrical, Langevin, and energy polarization terms. This is in contrast to valence band where the Peierls-Landau dominates while other contributions are suppressed.

This example illustrates that: (1) the geometrical magnetic susceptibility is an important contribution, especially around the band gap; (2) different terms in our classification dominate over different energy ranges; (3) it is possible to enhance the paramagnetic susceptibility by breaking the particle-hole symmetry. In addition, we note that for a generic two band model $\hat{h}=h_0+\bm h\cdot \bm \sigma$, $G_{n0}=-\bm B\cdot (\bm \partial h_0\times \bm A_{n0})$ is finite when particle-hole symmetry is broken. From Eq.(\ref{sus}), we find that only the Van Vleck paramagnetic susceptibility depends quadratically on $\bm \partial h_0$, while all other terms are either independent or only have linear dependence. This implies that Van Vleck susceptibility can in principle dominate for large $|\bm \partial h_0|$, which also leads to a strong paramagnetic response.

\section{Example II: Spin-1 continuum model}

In the previous example, we show that the single band properties contribute greatly to the orbital susceptibility, especially the geometrical magnetic susceptibility. As a second example, we consider the gapped spin-1 continuum model which is a three band system and hence has an enhanced interband Van Vleck paramagnetic susceptibility. We then compare the single band and interband contributions. 

The model Hamiltonian is $\hat{h}=v(k_1s_1+k_2 s_2)+\Delta s_3$, where $s_i$ are standard spin-1 matrices:\cite{Bercious2009}
\begin{equation}\label{spin}
s_1={1\over \sqrt{2}}
\begin{pmatrix}
0 && 1 && 0\\
1 && 0 && 1\\
0 && 1 && 0
\end{pmatrix}
\,,s_2={1\over \sqrt{2}}
\begin{pmatrix}
0 && -i && 0 \\
i && 0 && -i \\
0 && i && 0
\end{pmatrix}
\,,
\end{equation}
and $s_3$ is diagonal with entries $1$, $0$ and $-1$.

 This model describes the low energy behaviour of the dice lattice.\cite{Raoux2014} The band dispersions are $E=n\sqrt{\Delta^2+v^2k^2}$, with $n=-1, 0, 1$. If the Fermi energy falls in the lowest band, compared with corresponding magnetic susceptibilities in Eq.(\ref{Dirac}) for the spin-${1\over 2}$ model, Pauli and Peierls-Landau magnetic susceptibilities are the same, and the geometrical magnetic susceptibility doubles:
\begin{equation}\label{spinone1}
\chi_\text{P}={e^2\Delta^2 v^2\over 8\pi |\mu|^3},\; \chi_\text{L}=-{\chi_\text{P}\over 3},\; \chi_\text{Geom}=-{e^2\Delta^2 v^2\over 6 \pi |\mu|^3}.
\end{equation}
The Van-Vleck and $k$-space energy polarization magnetic susceptibilities are nontrivial and read :
\begin{equation}\label{spinone}
\chi_{\text{VV}}={e^2v^2\over 8\pi|\mu|}\left(1-{\Delta^2\over 3\mu^2}\right)\,,\chi_{\text{polar}}=-{e^2v^2\over 8\pi|\mu|}\left(1-{\Delta^2\over \mu^2}\right) \,.
\end{equation}

Compared with the spin-${1/2}$ continuum model, the total orbital susceptibility still vanishes if the Fermi energy falls in the band, but in the band gap the Van Vleck paramagnetic susceptibility can compete with the geometrical magnetic susceptibility: it is half the geometrical magnetic susceptibility with an opposite sign. There is another interesting feature about this model due to the flat band at zero energy which has a vanishing Berry curvature but nonvanishing magnetic moment. This interesting geometry yields a paramagnetic peak at zero temperature due to the Pauli paramagnetism: $\chi_{\text{flat}}=e^2v^2/(4\pi)\delta(\mu)$. Unlike the tight binding honeycomb model discussed previously, this Pauli paramagnetism is not compensated by the geometrical magnetic susceptibility since $\bm{\mathit\Omega}$ vanishes for the flat band. Given that this paramagnetic peak has a bigger magnitude than the geometrical magnetic susceptibility, this system will exhibit a paramagnetic response at low temperature when Fermi energy falls in the gap, in contrast to the diamagenetic response purely due to the geometrical magentism as discussed in the model in previous section. At the limit of vanishing band gap, the total paramagnetic susceptibility is the same as obtained in Ref.\onlinecite{Raoux2014}.

This example illustrates that: (1) generally if more bands are involved in the calculation of the orbital susceptibility, the interband effect (Van Vleck magnetic susceptibility) is enhanced, while the geometrical contribution is still prominant; (2) the total susceptibility is greatly affected by band geometries.

\section{Beyond minimal coupling}
Up to now, we are working under the minimal coupling assumption, i.e. the magnetic field enters the Hamiltonian only through the vector potential. This assumption is exact for the Dirac Hamiltonian with a periodic potential,  which yields the complete Hilbert space for electrons in solids. Therefore, our expression of the orbital magnetic susceptibility is also complete in this situation. However, for any effective Hamiltonians defined in a subspace of the complete Hilbert space, such as the Schr$\ddot{\rm o}$dinger Hamiltonian, the Pauli Hamiltonian with the spin-orbit coupling, and the effective Hamiltonian for some specific bands, the magnetic fields may have other effects. Our theory of the orbital magnetic susceptiblity can be easily extended to incorporate this situation.

Based on the Foldy-Wouthuysen transformation in solid state context, the Hamiltonian in a uniform magnetic field beyond minimal coupling assumption is :\cite{FW1950, Blount1962b, Chang2008} 
\begin{align}\label{fw}
\hat{H}&=\hat{H}_0(\hat{\bm p}+{1\over 2}\bm B\times \hat{\bm q},\bm q)-\bm B\cdot \hat{\bm \mu}(\hat{\bm p}+{1\over 2}\bm B\times \hat{\bm q})\notag\\
&\quad+B_i B_j \hat{h}_{ij}(\hat{\bm p}+{1\over 2}\bm B\times \hat{\bm q})\,,
\end{align}
where $\hat{\bm \mu}$ and $\hat{h}_{ij}$ are appropriate matrix operators as functions of $(\bm p+{1\over 2}\bm B\times \hat{\bm q})$. An example for $\hat{\bm \mu}$ is that if one derives the Pauli Hamiltonian from the Dirac Hamiltonian, one finds $\hat{\bm \mu}$ is the magnetic moment for spins: \cite{Blount1962b}
\begin{equation}\label{spinmu}
\hat{\bm \mu}={c^2 e\hbar\bm \sigma \over 2 \varepsilon_0}\,,
\end{equation}
where $\varepsilon_0$ is the electron band dispersion, and $\bm \sigma$ is the Pauli matrices for electron spins. The expression for $\hat{h}$ is also given in Ref.\onlinecite{Blount1962b}. 

We can expand the new Hamiltonian in Eq.(\ref{fw}) around the center of mass position $\bm r_c$. Compared with the same expansion for $\hat{H}_0$ as given in Sect.II, we find that the local Hamiltonian $\hat{H}_c=\hat{H}_0(\bm p+{1\over 2}\bm B\times \bm r_c,\bm q)$ is unchanged; the first order correction is
\begin{equation}\label{newh1}
\hat{H}^\prime=-{1\over 2}\bm B\cdot [\hat{\bm V}\times (\hat{\bm q}-\bm r
 _c)]-\bm B\cdot \hat{\bm \mu}(\bm p+{1\over 2}\bm B\times \bm r_c)\,,
\end{equation} 
and it receives a new term from $\hat{\bm \mu}$; the second order correction is
\begin{align}\label{newh2}
 \hat{H}^{\prime\prime}&={1\over 8}\mathit\Gamma_{ij} [\bm B\times (\hat{\bm q}-\bm r_c)]_i [\bm B\times (\hat{\bm q}-\bm r_c)]_j\notag\\
&+\left\{\left[{1\over 2}\bm B\times (\hat{\bm q}-\bm r_c)\right]\cdot \bm \partial\right\} \hat{\bm \mu}(\bm p+{1\over 2}\bm B\times \bm r_c) \cdot \bm B\notag\\
&+B_iB_j\hat{h}_{ij}(\bm p+{1\over 2}\bm B\times \bm r_c)\,.
\end{align} 

By following the same precedure as in Sect II-IV, we can derive the orbital magnetic susceptibility for the Hamiltonian in Eq.(\ref{fw}). The result is similar to Eq.(\ref{sus}), with the following modification:
(1) the second term in Eq.(\ref{newh1}) is of first order with respect to the magnetic field, and its influence to the susceptibility is through $\bm m$ and $\bm G_{n0}$: $\bm m \rightarrow \bm m+\langle 0|\hat{\bm \mu}|0\rangle$ and $G_{n0}\rightarrow G_{n0}+\bm B\cdot \langle n|\hat{\bm \mu}|0\rangle$; consequently, the Pauli magnetism in Eq.(\ref{sus}) will contain the contribution from $\hat{\bm \mu}$; the geometrical magnetic susceptibility and Van Vleck magnetic susceptibility are also affected; (2) the second and third term in Eq.(\ref{newh2}) is of second order with respect to the magnetic field and its expectation value directly adds to the second order wave-packet energy and modifies the Langevin magnetic susceptibility:
\begin{align}
g_{\text Langevin}&= -{f\over 8} \epsilon_{s i k} \epsilon_{t j \ell}B_{s} B_{t}g_{ij}\alpha_{k\ell}+fB_iB_j \langle u_0 | \hat{h}_{ij}|u_0\rangle\notag\\
&+{f\over 4}[(\bm B\times \bm A_{0n})_i\langle u_n|\partial_i (\bm B\cdot \hat{\bm \mu})|u_0\rangle+c.c.]\,.
\end{align}

\section{Summary}

In summary, we derive a compact gauge-invariant expression of the Bloch wave-packet energy correct to second order in external magnetic field, which fully incorporates the first order correction of the wave-packet from vertical and horizontal mixings. Based on this, we obtain a complete and compact formula for the orbital magnetic susceptibility, with important advantage that each term is gauge invariant and has clear physical meanings. We demonstrate that other than the familiar Pauli and Peierls-Landau magnetic susceptibilities, the orbital susceptibility in solids also consists of the Van Vleck paramagnetic susceptibility and Langevin magnetic susceptibility, which reduce to their counterparts in atomic physics in the limit of atomic insulators. More importantly, we identify two new contributions: the geometrical magnetic susceptibility derived from the Berry curvature and the  quantum metric which can dominate in a small energy gap, and the $k$-space energy polarization magnetic susceptibility, which competes with Peierls-Landau and Pauli magnetic susceptibility on a Fermi surface. We illustrate that the Pauli, Peierls-Landau and geometrical magnetic susceptibility depend solely on band geometrical quantities and affect the orbital susceptibility greatly, as shown in two examples.

{\it Acknowledgment.}---We acknowledge valuable discussions with D. L. Deng, Z. Qiao, H. Chen, J. Zhou, X. Li, R. Cheng, and L. Zhang. We thank  Shan Guan for helping with the figures. We thank J.-N. Fuchs, A.Raoux, F. Pi\'{e}chon and G. Montambaux for pointing out a numerical mistake in our previous manuscript. QN is supported by NBRPC (No.2012CB921300 and No.2013CB921900), and NSFC (No.91121004). SAY is supported by SUTD-SRG-EPD-2013062. YG is supported by DOE (DE-FG03-02ER45958, Division of Materials Science and Engineering) and Welch Foundation (F-1255).

\appendix
\section{derivation of the wave-packet energy}
The wave-packet energy has several parts. We first consider the contribution from $\hat{H}_c$ first.
\begin{align}\label{lc}
\langle \Psi|\hat{H}_c|\Psi\rangle=\int d\bm p C_0^\star C_0 \varepsilon_0 +\sum_{n\neq 0}\int d\bm p C_n^\star C_n \varepsilon_n\,.
\end{align}

There is an important correction to Eq.(\ref{lc}): when the mixing of Bloch states $C_n$ is taken into account, the normalization of the wave-packet is modified:
\begin{equation}\label{norm}
\langle \Psi|\Psi\rangle=\int \left(|C_0|^2+\sum_{n\neq 0}|C_n|^2\right)) d^3 \bm p\,.
\end{equation}

To make the Euler-Lagrangian method valid, the wave-packet must be normalized, at least up to second order for our purpose. Therefore, $C_0$ must have a second order correction: $C_0\rightarrow (1+\delta)C_0$, with $\delta=-{1\over 2} \sum_{n\neq 0}\int |C_n|^2 d^3p$. 

This correction $\delta$ contributes to the energy in Eq.(\ref{lc}): 
\begin{align}\label{e2}
\langle \Psi|\hat{H}_c|\Psi\rangle&= \varepsilon_0-\sum_{n\neq 0} \int d\bm p C_n^\star C_n (\varepsilon_0 -\varepsilon_n)+\delta \varepsilon_c\,.
\end{align}
Here, the term $\delta \varepsilon_c$ arises due to the horizontal mixing in the coefficient $C_n$. Compared with the renormalization condition in Eq.(\ref{norm}), the integration in Eq.(\ref{lc}) has an additional energy factor $\varepsilon_0$ in the integrand, which leads to the additional term $\delta \varepsilon_c$ due to the derivative of $C_0$ involved in the horizontal mixing. However, this $\delta \varepsilon_c$ is cancelled by the contribution from the dynamical part of the Langrangian:
\begin{equation}\label{dl}
\delta L_{dyn}=\int d\bm p C_0^\star \delta  i\partial_t C_0 -\sum_{n\neq 0}\int d\bm p C_n^\star i\partial_t C_n\,.
\end{equation}

We only need to keep the term up to the second order in $\bm B$:
\begin{align}\label{dcn}
i\partial_t C_n&={G_{n0} \over \varepsilon_0 -\varepsilon_n} i\partial_t C_0 +{i\over 2} (\bm B\times \bm A_{n0})\cdot (i \bm v_0-i\dot{\bm r}_c) C_0\notag\\
&\quad + \varepsilon_0 {i\over 2}(\bm B\times \bm A_{n0})\cdot (\hat{\bm D}-\bm r_c) C_0\,.
\end{align}
Plug Eq.(\ref{dcn}) into Eq.(\ref{dl}), and we have
\begin{equation}\label{e3}
\delta L_{dyn}= -\delta \varepsilon_c +\sum_{n\neq 0}{1\over 8} \alpha_{ij} (\bm B\times \bm A_{0n})_i (\bm B\times \bm A_{n0})_j\,,
\end{equation}
where $\alpha_{ij}=\partial_{ij} \varepsilon_0$ is the inverse effective mass tensor.

Then we calculate the contribution to the wave-packet energy from the gradient correction $\hat{H}^\prime$:
\begin{align}\label{epsilon1}
\langle \Psi|\hat{H}^\prime|\Psi\rangle
&=-\bm B\cdot \bm m+\bm B\cdot (\bm v_0\times \bm a_0^\prime)\notag\\
&+2\sum_{n\neq 0}{G_{0n}G_{n0}\over \varepsilon_{0}-\varepsilon_{n}}-\sum_{n\neq 0}{1\over 4} {\alpha}_{ij} (\bm B\times \bm A_{0n})_i (\bm B\times \bm A_{n0})_j\notag\\
&+\sum_{n\neq 0}{1\over 2}\partial_i [ (\bm B\times \bm A_{0n})_i G_{n0}+c.c.]\notag\\
&+\sum_{n\neq 0}{1\over 4} \int d\bm p \{ [\bm B\times (\hat{\bm D}-\bm r_{c})]_i^\star C_0^\star [\bm B\times (\hat{\bm D}-\bm r_{c})]_jC_0\notag\\
&\quad\quad\quad  [-i(V_i)_{0n}(A_j)_{n0}]+c.c.\}\,,
\end{align}
where $\bm m$ is the orbital magnetic moment: $\bm m=-{1\over 2} {\rm Im}\langle \bm \partial u_0|\times (\varepsilon_0-\hat{H}_c) |\bm \partial u_0\rangle$, and $\bm a_0^\prime$ is the positional shift: \cite{Yang2014}
\begin{equation}
\bm a_0^\prime=\sum_{n\neq 0} {G_{0n}\bm A_{n0}\over \varepsilon_0-\varepsilon_n}+{1\over 4}\partial_i[(\bm B\times \bm A_{0n})_i\bm A_{n0}]+c.c.\,.
\end{equation}

The remaining contribution to the wave-packet energy is from the second order correction $\hat{H}^{\prime\prime}$:
\begin{align}\label{H21}
\langle \Psi | \hat{H}^{\prime\prime}|\Psi\rangle
&=-{1\over 8}\langle 0|\Gamma_{ij} |0\rangle (\bm B\times \bm r_c)_i(\bm B\times \bm r_c)_j\notag\\
&-\sum_{n\neq 0}{1\over 8} [(\bm B\times \bm A_{0n})_i (\Gamma_{ij})_{n0} (\bm B\times \bm r_c)_j+c.c.]\notag\\
&+{1\over 8} \int d\bm p (\bm B\times \bm \partial |C_0|)_i (\bm B\times \bm \partial |C_0|)_j \langle 0| \Gamma_{ij}|0\rangle \notag\\
&+{1\over 8} \langle 0|\Gamma_{ij}|0\rangle (\bm B\times \bm r_c)_i (\bm B\times \bm r_c)_j\notag\\
&+\sum_{n\neq 0}{1\over 8} [(\bm B\times \bm A_{0n})_i (\Gamma_{ij})_{n0} (\bm B\times \bm r_c)_j+c.c.]\notag\\
&+\sum_{n\neq 0}{1\over 16} (\bm B\times i\bm \partial)_i [(\Gamma_{ij})_{0n} (\bm B\times \bm A_{n0})_j]-(i\leftrightarrow j)\notag\\
&+{1\over 8} \sum_{(m,n)\neq 0} (\bm B\times \bm A_{0m})_i (\Gamma_{ij})_{mn} (\bm B\times \bm A_{n0})_j\,.
\end{align}
After some cancellations, we have
\begin{align}\label{H2}
\langle \Psi | \hat{H}^{\prime\prime}|\Psi\rangle
&=-{1\over 16} (\bm B\times \bm \partial)_i (\bm B\times \bm \partial)_j \langle 0|\mathit{\Gamma}_{ij}|0\rangle\notag\\
&+{1\over 8} \sum_{(m,n)\neq 0} (\bm B\times \bm A_{0m})_i (\mathit{\Gamma}_{ij})_{mn} (\bm B\times \bm A_{n0})_j\,,
\end{align}
where $(\mathit{\Gamma}_{ij})_{mn}=\langle u_m|\mathit{\Gamma}_{ij}|u_n\rangle$. For relativistic Dirac Hamiltonian, $L^{\prime\prime}$ simply vanishes. For nonrelativistic Schr$\ddot{\rm o}$dinger and Pauli Hamiltonian, $\mathit{\Gamma}_{ij}$ is inverse mass times the identity matrix, and Eq.(\ref{H2}) reduces to a compact form: 
\begin{equation}\label{H2sim}
\langle \Psi | \hat{H}^{\prime\prime}|\Psi\rangle={1\over 8m} (B^2 g_{ii}-B_i g_{ij} B_j)\,,
\end{equation}
where $g_{ij}={\rm Re}\langle \partial_i u_0|\partial_j u_0\rangle-a_i a_j$ is the quantum metric of $k$-space.\cite{anandan1990, Neupert2013}

By combining Eq.(\ref{e2}),(\ref{e3}),(\ref{epsilon1}),(\ref{H2sim}) and the relation between effective mass tensor $\alpha_{ij}$ and the Hessian matrix $\mathit{\Gamma}_{ij}$ of $\hat{H}_c$:
\begin{equation}
\langle 0|\mathit{\Gamma}_{ij}|0\rangle=\alpha_{ij}+\sum_{n\neq 0}[-i (A_i)_{0n} (V_j)_{n0}+c.c.]\,,
\end{equation}
the wave-packet energy can be put in a compact form (assume $\mathit{\Gamma}_{ij}=\delta_{ij}/m$):
\begin{align}\label{ap_energy}
\tilde{\varepsilon}=\,&\varepsilon_0 -\bm B\cdot \bm m\notag\\
&+{1\over 4} (\bm B\cdot \bm{\mathit\Omega}) (\bm B\cdot \bm m)-{1\over 8} \epsilon_{s i k} \epsilon_{t j \ell}B_{s} B_{t}g_{ij} \alpha_{k\ell} \notag\\
& -\bm B\cdot (\bm a_0^\prime \times \bm v_0)+\nabla\cdot\bm P_\text{E}\notag\\
&+\sum_{n\neq 0}{G_{0n} G_{n0}\over \varepsilon_{0}-\varepsilon_{n}}+{1\over 8m} (B^2 g_{ii}-B_i g_{ij} B_j).
\end{align}
Various quantities in Eq.(\ref{ap_energy}) have been explained in the main text. Note that originally, physical quantities such as $|u_0\rangle$ and $|u_n\rangle$ in Eq.(\ref{bloch}) and (\ref{wavepacket}), $\bm A_{0n}$, $G_{n0}$, $\varepsilon_0$ and $\varepsilon_n$ in Eq.(\ref{bloch}), (\ref{Cn}), (\ref{lc}) and (\ref{dcn}) are all functions of $\bm p+{1\over 2}\bm B\times \bm r_c$. However, since $|C_0|^2$ is a delta function localized around $\bm p_c$, after intergration with $|C_0|^2$, those physical quantities in the final results in Eq.(\ref{energy}), (\ref{epsilon1}), (\ref{H2}) and (\ref{ap_energy}) are functions of the physical momentum $\bm k_c=\bm p_c+{1\over 2}\bm B\times \bm r_c$.

\section{k-space Energy Polarization}
The $k$-space polarization energy can be expressed in terms of the single band Bloch states $|u_0\rangle$:
\begin{equation}\label{pol}
{1\over 4}\epsilon_{sik}\epsilon_{tj\ell}B_s B_t\partial_k [\langle \hat{D}_i u_0|(\hat{ V}+v_0)_\ell |\hat{D}_j u_0\rangle+c.c.]\,,
\end{equation}
where $\hat{D}_i=\partial_i+i (a_0)_i$ is the covariant derivative acting on the Bloch state $|u_0\rangle$. In Eq.(\ref{pol}), the first term is from the gradient correction to $\hat{H}_c$, and the second term is from the nonadiabatic correction. To understand the geometric meaning of Eq.(\ref{pol}), we consider the quadrupole moment of the velocity operator under the wave-packet $|\Psi\rangle$:
\begin{align}\label{quad}
&\langle \Psi| (\bm q-\bm r_c)_i \hat{V}_k (\bm q-\bm r_c)_j +c.c.|\Psi\rangle\notag\\
&=\int  d\bm p [(i\bm \partial+\bm a_0-\bm r_c)_i^\star C_0^\star] [(i\bm \partial+\bm a_0-\bm r_c)_j C_0] (\bm v_0)_k+c.c.\notag\\
&+\int d\bm p [(i\bm \partial+\bm a_0-\bm r_c)_i^\star C_0^\star] C_0 \langle u_0|\hat{V}_k |\hat{D}_j u_0\rangle+c.c.\notag\\
&+\int  d\bm p  C_0^\star [(i\bm \partial+\bm a_0-\bm r_c)_j C_0] \langle \hat{D}_i u_0|\hat{V}_k|u_0\rangle+c.c.\notag\\
&+\int d\bm p C_0^\star C_0 \langle \hat{D}_iu_0|\hat{V}_k|\hat{D}_j u_0\rangle+c.c.\,.
\end{align}

The $k$-space energy polarization is connected to the quadrupole moment of the velocity operator:
\begin{align} 
&\quad\; {1\over 2}(\hat{H}^\prime \delta \bm p+c.c.)\notag\\
&={1\over 4}\bm B\times (\hat{H}^\prime (\hat{\bm q}-\bm r_c)+c.c.)\notag\\
&={1\over 4}\hat{e}_\ell\epsilon_{si\ell}\epsilon_{tjk}B_s B_t (\hat{\bm q}-\bm r_c)_i \hat{V}_k (\hat{\bm q}-\bm r_c)_j+c.c.\,.
\end{align}

The first term in Eq.(\ref{quad}) vanishes. Meanwhile, notice that $\delta \bm p$ in the definition of the energy polarization indicates that the horizontal mixing in the wave-packet is used. Since the energy polarization combines the horizontal and vertical mixing, we should not take the intraband part of $\hat{\bm q}-\bm r_c$ in $\hat{H}^\prime$ in the Bloch representation which represents the horizontal mixing in $\hat{H}^\prime$. Therefore, the remaining two contributions from Eq.(\ref{quad}) yield:
\begin{align}
&\quad\;{1\over 2}(\hat{H}^\prime \delta \bm p+c.c.)\notag\\
&={1\over 4}\hat{e}_\ell\epsilon_{si\ell}\epsilon_{tjk}B_s B_t {-i\over 2} \partial_i [\langle u_0 |\hat{V}_k |\hat{D}_ju_0\rangle+c.c.]\notag\\
&+{1\over 4}\hat{e}_\ell \epsilon_{si\ell}\epsilon_{tjk}B_s B_t [\langle \hat{D}_i u_0|\hat{ V}_k |\hat{D}_j u_0\rangle+c.c.]\,,
\end{align}

The resulting $k$-space polarization energy is
\begin{align}
&\quad\;{1\over 2}\bm \partial \cdot (\hat{H}^\prime \delta \bm p+c.c.)\notag\\
&={1\over 4} \epsilon_{si\ell}\epsilon_{tjk}B_s B_t {-i\over 2} \partial_\ell\partial_i [\langle u_0 |\hat{V}_k |\hat{D}_ju_0\rangle+c.c.]\notag\\
&+{1\over 4} \epsilon_{si\ell}\epsilon_{tjk}B_s B_t \partial_\ell [\langle \hat{D}_i u_0|\hat{ V}_k |\hat{D}_j u_0\rangle+c.c.]\notag\\
&={1\over 4} \epsilon_{si\ell}\epsilon_{tjk}B_s B_t \partial_\ell [\langle \hat{D}_i u_0|\hat{ V}_k |\hat{D}_j u_0\rangle+c.c.]
\end{align}

This yields the first term in Eq.(\ref{pol}). To obtain the second term, notice that the Bloch state $|u_0\rangle$ depends on time through $\bm r_c$, which yields a first order Hamiltonian:
\begin{equation}
\langle u_n |i\partial_t |u_0\rangle={1\over 2}\bm B\times \bm v_0 \cdot \langle u_n |i\bm \partial_{\bm p}|u_0\rangle={1\over 2}\bm B\times \bm v_0\cdot \bm A_{n0}\,.
\end{equation}
Through similar derivations for the polarization energy from $\hat{H}^\prime$, it is straightforward to derive the second term in Eq.(\ref{pol}).

Finally, it is interesting that through similar derivations in Eq.(\ref{quad}), we find that the quantum metric $g_{ij}$ is the intrinsic expansion of the wave-packet:
\begin{align}\label{metric}
&\quad\;\langle \Psi |(\hat{\bm q}-\bm r_c)_i (\hat{\bm q}-\bm r_c)_j+c.c. |\Psi\rangle\notag\\
&=2\int d\bm p (\partial_i|C_0|)(\partial_j|C_0|)+2\int d\bm p |C_0|^2 g_{ij}\,.
\end{align}
In Eq.(\ref{metric}), the first term on the right hand side is the extrinsic expansion, which is determined by the shape of the wave-packet. From the semiclassical point of view, the particle simultaneously has the momentum and position as dynamical variables, and the extrinsic expansion of $|C_0|$ is ignored. However, the second term in Eq.(\ref{metric}) reflects the intrinsic expansion of the wave packet, due to the projection of the whole Hilbert space onto a single band Hilbert space.

\section{susceptibilities in the atomic insulator limit}

We will discussion the Van Vleck paramagentism in the Wannier function representation first.
\begin{widetext}
\begin{align} \label{wanVV}
&\quad\int {d^3k_c\over 8\pi^3} \sum_{n}f(\varepsilon_0(\bm k_c)){G_{0n}G_{n0}\over \varepsilon_0-\varepsilon_n}\notag\\
&=\int {d^3k_c\over 8\pi^3} \sum_n {f(\varepsilon_0(\bm k_c))\over \varepsilon_0-\varepsilon_n}{1\over N}\big |\sum_{\bm R_1,\bm R_2}{1\over 2} e^{i\bm k_c\cdot (\bm R_1-\bm R_2)}\bm B\cdot\notag\\
&   [\langle W_n(\bm R_2)|\hat{\bm V}\times (\bm r-\bm R_1)|W_0(\bm R_1)\rangle
-{1\over N}\sum_{\bm R_3,\bm R_4}e^{i\bm k_c\cdot (\bm R_3-\bm R_4)} \langle W_n(\bm R_2)|\hat{\bm V}|W_0(\bm R_1)\rangle \times\langle W_0(\bm R_4) |(\bm r-\bm R_3)|W_0(\bm R_3)\rangle] \big |^2\notag\\
&=\int {d^3k_c\over 8\pi^3} \sum_n {f(\varepsilon_0(\bm k_c))\over \varepsilon_0-\varepsilon_n}{1\over N}\big |\sum_{\bm R_1}{1\over 2}\bm B\cdot \notag\\
&[\langle W_n(\bm R_1)|\hat{\bm V}\times (\bm r-\bm R_1) |W_0(\bm R_1)\rangle-{1\over N} \sum_{\bm R_3}\langle W_n(\bm R_1)|\hat{\bm V}|W_0(\bm R_1)\rangle \times\langle W_0(\bm R_3) |(\bm r-\bm R_3)|W_0(\bm R_3)\rangle]|^2\notag\\
&= {N\over V}\sum_n{f(\varepsilon_0)\over \varepsilon_0-\varepsilon_n}\big | \langle W_n(\bm R)|{1\over 2}\bm B\cdot[\hat{\bm V}\times (\bm r-\bm R)] |W_0(\bm R)\rangle|^2\,.
\end{align}
\end{widetext}
In Eq.(\ref{wanVV}), the first equality is simply the expansion of the Van Vleck paramangetism for a single band $\varepsilon_0$ in terms of the Wannier function. The second equality is derived by taking the atomic insulator limit and setting all inter-lattice-site hopping to zero: $\langle W_n(\bm R_2)|\hat{\bm V}\times (\bm r-\bm R_1)|W_0(\bm R_1)\rangle\propto \delta_{\bm R_1,\bm R_2}$, $\langle W_n(\bm R_2)|\hat{\bm V}|W_0(\bm R_1)\rangle\propto \delta_{\bm R_1,\bm R_2}$ and $\langle W_0(\bm R_4)|(\bm r-\bm R_3)|W_0(\bm R_3)\rangle\propto \delta_{\bm R_3,\bm R_4}$. The last equality accounts for the fact that for atomic insulators, we only have flat band $\varepsilon_0$ with no $\bm k_c$ dependence, which coincides with the atomic energy level. Meanwhile, $\langle W_0(\bm R_3)|(\bm r-\bm R_3)|W_0(\bm R_3)\rangle=0$, since the electron is bound with each atom and its position expectation should coincide with the atom position. We also use the fact that $\langle W_n(\bm R)|{1\over 2}\bm B\cdot[\hat{\bm V}\times (\bm r-\bm R)] |W_0(\bm R)\rangle$ are identical for all lattice sites. 

Now we calculate the Langevin magnetic free energy:
\begin{widetext}
\begin{align} \label{wanLan}
&\quad\int {d^3k_c\over 8\pi^3} {f(\varepsilon_0(\bm k_c))\over 8m} (B^2 g_{ii}-B_i g_{ij} B_j)\notag\\
&=\int {d^3k_c\over 8\pi^3}  {f(\varepsilon_0(\bm k_c))\over 8m}\epsilon_{\ell si}\epsilon_{\ell tj} B_s B_t{1\over N}\sum_{\bm R_1,\bm R_2} e^{i\bm k_c\cdot (\bm R_1-\bm R_2)}\notag\\
&  \left( \langle W_0(\bm R_2)|(\bm r-\bm R_2)_i (\bm r-\bm R_1)_j|W_0(\bm R_1)\rangle
-{1\over N}\sum_{\bm R_3,\bm R_4}e^{i\bm k_c\cdot (\bm R_3-\bm R_4)} \langle W_0(\bm R_2)|(\bm r-\bm R_1)_j|W_0(\bm R_1)\rangle \langle W_0(\bm R_4) |(\bm r-\bm R_3)_i|W_0(\bm R_3)\rangle\right)\notag\\
&=\int {d^3k_c\over 8\pi^3}  {f(\varepsilon_0(\bm k_c))\over 8m}\epsilon_{\ell si}\epsilon_{\ell tj} B_s B_t{1\over N}\sum_{\bm R_1} \notag\\
&  \left( \langle W_0(\bm R_1)|(\bm r-\bm R_1)_i (\bm r-\bm R_1)_j|W_0(\bm R_1)\rangle
-{1\over N}\sum_{\bm R_3} \langle W_0(\bm R_1)|(\bm r-\bm R_1)_j|W_0(\bm R_1)\rangle \langle W_0(\bm R_3) |(\bm r-\bm R_3)_i|W_0(\bm R_3)\rangle\right)\notag\\
&={N\over V}{f(\varepsilon_0)\over 8m} \langle W_0(\bm R)||\bm B\times (\bm r-\bm R)|^2|W_0(\bm R)\rangle\,.
\end{align}
\end{widetext}
This reduce to the familiar Langevin magentism for atomic systems.

For the geometrical magentism, we will first calculate the first term:
\begin{widetext}
\begin{align} \label{wangeo1}
&\quad -{3\over 4}\int {d^3k_c\over 8\pi^3} f(\varepsilon_0(\bm k_c)) (\bm B\cdot \bm{\mathit\Omega})(\bm B\cdot \bm m)\notag\\
&={3\over 8} \int {d^3k_c\over 8\pi^3} f(\varepsilon_0(\bm k_c)){1\over N^2}\sum_{\bm R_1,\bm R_2,\bm R_3,\bm R_4} e^{i\bm k_c \cdot (\bm R_1+\bm R_3-\bm R_2-\bm R_4)}\notag\\
&  \left( \langle W_0(\bm R_2)|[(\bm r-\bm R_2)\times (\bm r-\bm R_1)]\cdot \bm B|W_0(\bm R_1)\rangle \langle W_0(\bm R_4) |[\bm B\times(\bm r-\bm R_4)]\cdot [(\varepsilon_0-\hat{H}_c)(\bm r-\bm R_3)]|W_0(\bm R_3)\rangle\right)\notag\\
&={3\over 8} {f(\varepsilon_0)\over N^2} \sum_{\bm R_1,\bm R_2,\bm R_3,\bm R_4} \delta_{(\bm R_1+\bm R_3-\bm R_2-\bm R_4),0}\notag\\
&  \left( \langle W_0(\bm R_2)|[(\bm r-\bm R_2)\times (\bm r-\bm R_1)]\cdot \bm B|W_0(\bm R_1)\rangle \langle W_0(\bm R_4) |[\bm B\times(\bm r-\bm R_4)]\cdot [(\varepsilon_0-\hat{H}_c)(\bm r-\bm R_3)]|W_0(\bm R_3)\rangle\right)\notag\\
\end{align}
\end{widetext}
In the above equation, the second equality uses the fact that in the atomic insulator limit, the band dispersion becomes flat. Notice that in the last line in Eq.(\ref{wangeo1}), $(\bm r-\bm R_2)\times (\bm r-\bm R_1)=0$ if $\bm R_1 =\bm R_2$. Therefore, to obtain a nonzero contribution, we must have $\bm R_1\neq \bm R_2$ and $\bm R_3 \neq \bm R_4$ due to the Kronecker delta function. This part of geometrical contribution contains only inter-lattice-hopping effect, and vanishes in the limit of atomic insulators. 

The second term in the geometrical magnetic free energy reads:
\begin{widetext}
\begin{align} \label{wangeo2}
&\quad -{1\over 8}\int {d^3 k_c\over 8\pi^3}f(\varepsilon_0(\bm k_c))\epsilon_{s i k} \epsilon_{t j \ell}B_{s} B_{t}g_{ij}\alpha_{k\ell}\notag\\
&={1\over 8} \int {d^3k_c\over 8\pi^3} f(\varepsilon_0(\bm k_c))\epsilon_{s i k} \epsilon_{t j \ell}B_{s} B_{t}{1\over N^2}\sum_{\bm R_1,\bm R_2,\bm R_3,\bm R_4} e^{i\bm k_c \cdot (\bm R_1+\bm R_3-\bm R_2-\bm R_4)}(\bm R_3-\bm R_4)_k(\bm R_3-\bm R_4)_\ell\langle W_0(\bm R_4)|\hat{H}_c|W_0(\bm R_3)\rangle\notag\\
&  \left( \langle W_0(\bm R_2)|(\bm r-\bm R_2)_i (\bm r-\bm R_1)_j|W_0(\bm R_1)\rangle
-{1\over N}\sum_{\bm R_3,\bm R_4}e^{i\bm k_c\cdot (\bm R_5-\bm R_6)} \langle W_0(\bm R_2)|(\bm r-\bm R_1)_j|W_0(\bm R_1)\rangle \langle W_0(\bm R_6) |(\bm r-\bm R_5)_i|W_0(\bm R_5)\rangle\right)\notag\\
&={1\over 8} {f(\varepsilon_0)\over N^2} \epsilon_{s i k} \epsilon_{t j \ell}B_{s} B_{t}\sum_{\bm R_1,\bm R_2,\bm R_3,\bm R_4} \delta_{(\bm R_1+\bm R_3-\bm R_2-\bm R_4),0}\notag\\
& (\bm R_3-\bm R_4)_k(\bm R_3-\bm R_4)_\ell\langle W_0(\bm R_4)|\hat{H}_c|W_0(\bm R_3)\rangle\langle W_0(\bm R_2)|(\bm r-\bm R_2)_i (\bm r-\bm R_1)_j|W_0(\bm R_1)\rangle\,.
\end{align}
\end{widetext}
Due to the prefactor $(\bm R_3-\bm R_4)$, it is obvious that this remaining contribution to the geometrical magnetic free energy also contains only inter-lattice-hopping effect, and vanishes in the limit of atomic insulators.

\bibliographystyle{apsrev4-1}

\begin{thebibliography}{100}
\bibitem[1]{Mermin1976}N. W. Ashcroft and N. D. Mermin, Solid State Physics (Saunders, Philadelphia, 1976).
\bibitem[2]{Xiao2010} D. Xiao, M.-C. Chang, and Q. Niu, Rev. Mod. Phys. 82, 1959 (2010).
\bibitem[3]{Nagaosa2010}N. Nagaosa, J. Sinova, S. Onoda, A. H. MacDonald and N. P. Ong, Rev. Mod. Phys. 82, 1539 (2010).
\bibitem[4]{Kagan2008}Y. Kagan and L. A. Maksimov, Phys. Rev. Lett. 100, 145902 (2008).
\bibitem[5]{Zhang2010}L. Zhang, J. Ren, J.-S. Wang, and B. Li, Phys. Rev. Lett. 105, 225901 (2010).
\bibitem[6]{Qin2012}T. Qin, J. Zhou, and J. Shi, Phys. Rev. B 86, 104305 (2012).
\bibitem[7]{Onoda2006} S. Onoda, N. Sugimoto, and N. Nagaosa, Phys. Rev. Lett. 97, 126602 (2006).
\bibitem[8]{Shi2007} J. Shi, G. Vignale, D. Xiao and Q. Niu, Phys. Rev. Lett. 99, 197202 (2007).
\bibitem[9]{Adams1953} E. N. Adams, Phys. Rev. 89, 633 (1953).
\bibitem[10]{Hebborn1960} J. E. Hebborn and E. H. Sondheimer, J. Phys. Chem. Solids 13, 105 (1960).
\bibitem[11]{Kohn1955} J. M. Luttinger and W. Kohn, Phys. Rev. 97, 869 (1955).
\bibitem[12]{Roth1962} L. M. Roth, J. Phys. Chem. Solids 23, 433 (1962).
\bibitem[13]{Blount1962} E. I. Blount, Phys. Rev. 126, 1636 (1962).
\bibitem[14]{Wannier1964} G. H. Wannier and U. N. Upadhyaya, Phys. Rev. 136, A803 (1964).
\bibitem[15]{Misra1972} P. K. Misra and L. Kleinman, Phys. Rev. B. 5, 4581 (1972).
\bibitem[16]{Koshino2007} M. Koshino and T. Ando, Phys. Rev. B 76, 085425 (2007).
\bibitem[17]{Schober2012} G. A. H. Schober, H. Murakawa, M. S. Bahramy, R. Arita, Y. Kaneko, Y. Tokura, and N. Nagaosa, Phys. Rev. Lett. 108, 247208 (2012).
\bibitem[18]{Santos2011} G. G\'{o}mez-Santos and T. Stauber, Phys. Rev. Lett. 106, 045504 (2011).
\bibitem[19]{Raoux2014} A. Raoux, M. Morigi, J.-N. Fuchs, F. Pi\'{e}chon, and G. Montambaux, Phys. Rev. Lett. 112, 026402 (2014).
\bibitem[20]{Raoux2014b} A. Raoux, F. Pi\'{e}chon, J.-N. Fuchs, and G. Montambaux, arXiv:1411.5940.
\bibitem[21]{Fuku1971} H. Fukuyama, Prog. Theor. Phys. 45, 704 (1971).
\bibitem[22]{vignale1991} G. Vignale, Phys. Rev. Lett. 67, 358 (1991).
\bibitem[23]{Yang2014} Y. Gao, S. A. Yang and Q. Niu, Phys. Rev. Lett. 112, 166601 (2014).
\bibitem[24]{anandan1990} J. Anandan and Y. Aharonov, Phys. Rev. Lett. 65, 1697 (1990).
\bibitem[25]{Neupert2013} T. Neupert, C. Chamon, and C. Mudry, Phys. Rev. B 87, 245103 (2013).
\bibitem[26]{Van1993} R. D. King-Smith and D. Vanderbilt, Phys. Rev. B 47, 1651 (1993).
\bibitem[27]{Cai2013} T. Cai, S. A. Yang, X. Li, F. Zhang, J. Shi, W. Yao, and Q. Niu, Phys. Rev. B 88, 115140 (2013).
\bibitem[28]{Xiao2012} D. Xiao, G.-B. Liu, W. Feng, X. Xu and W. Yao, Phys. Rev. Lett. 108, 196802 (2012).
\bibitem[29]{NG2009} A. H. Castro Neto, F. Guinea, N. M. R. Peres, K. S. Novoselov and A. K. Geim, Rev. Mod. Phys. 81, 109 (2009).
\bibitem[30]{Zhang2011} X.-L. Qi and S.-C. Zhang, Rev. Mod. Phys. 83, 1057 (2011).
\bibitem[31]{Kane2010} M. Z. Hasan and C. L. Kane, Rev. Mod. Phys. 82, 3045 (2010).
\bibitem[32]{Xiao2007} D. Xiao, W. Yao, and Q. Niu, Phys. Rev. Lett. 99, 236809 (2007).
\bibitem[33]{Bercious2009} D. Bercious, D.F. Urban, H. Grabert, and W. H$\ddot{\rm a}$usler, Phys. Rev. A 80, 063603 (2009).
\bibitem[34]{FW1950}L. L. Foldy and S. A. Wouthuysen, Phys. Rev. 78, 29 (1950).
\bibitem[35]{Blount1962b} E. I. Blount, Phys. Rev. 128, 2454 (1962).
\bibitem[36]{Chang2008} M.-C. Chang and Q. Niu, J. Phys.: Condens. Matter 20, 193202 (2008).
\end{thebibliography}

\end{document}